\documentclass[a4paper]{jpconf}
\usepackage{amsmath}
\usepackage{latexsym}
\usepackage{graphicx}
\usepackage{color}
\usepackage{bm}
\usepackage{float}
\usepackage{bbold}
\usepackage{amsfonts,amssymb}
\usepackage[normalem]{ ulem }
\usepackage{soul}
\usepackage{verbatim}
\usepackage{scalerel}

\definecolor{Blue}{rgb}{0,0,1}
\definecolor{Red}{rgb}{1,0,0}

\newcommand{\N}[1]{{\mathcal N}_{#1}}
\usepackage{hyperref} %Hyrperref package: only for revtex version
\usepackage[all]{hypcap}

\begin{document}
\title{Parameter estimation via indefinite causal structures}

\author{Lorenzo M. Procopio}

\address{Weizmann Institute of Science, Rehovot 7610001, Israel}

\ead{lorenzo.procopio@weizmann.ac.il}

\begin{abstract}
Quantum Fisher information is the principal tool used to give the ultimate precision bound on the estimation of parameters for quantum channels. In this work, we present analytical expressions for the quantum Fisher information with three noisy channels for the case where the channels are in superposition of causal orders. We found that the quantum Fisher information  increases as the number of causal orders increases for certain combinations. We also show that certain combinations of causal orders attain higher precision on bounds than others for the same number of causal orders. Based on our results, we chose the best combinations of causal orders with three channels for probing schemes using indefinite causal structures.

\end{abstract}

\section{Introduction}

In the field of quantum metrology, the quantum channel identification problem attempts to estimate the true value of parameters from quantum channels \cite{fujiwara2005quantum}. The main task is to find the best strategy to efficiently determine the parameters of a channel with a minimum of resources while maintaining the desired precision. Initially, quantum entanglement was proposed as a strategy to estimate the isotropic depolarization parameter within a noisy channel \cite{fujiwara2005quantum}.  Other strategies, such as unentangled probes, have been also proposed for channel identification \cite{frey2011probing}. Quantum Fisher information is a metric used to learn about the value of parameters of quantum channels. To assess and compare different probing strategies, the quantum Fisher information is used as a figure of merit \cite{paris2009quantum}.  In all the above strategies, the order of the application of channels are in a definite causal order, in which the connections between channels are classically connected.

A new paradigm has recently been proposed in the field of quantum communications according to which the connections between quantum channels can be in a superposition of trajectories in space \cite{abbott2020communication} or time \cite{chiribella2019quantum}. By placing the quantum channels in a superposition of trajectories, one obtains an indefinite causal structure known as the quantum switch, which has been shown to be useful for new applications in quantum discrimination of channels \cite{chiribella2012perfect}, quantum computation \cite{Chiribella2013}, quantum communication complexity \cite{guerin2016exponential}, quantum metrology \cite{zhao2020quantum}, and quantum thermodynamics \cite{felce2020quantum}. These theoretical investigations have motived experimental demonstrations with two channels \cite{procopio2015experimental,rubino2017experimental,goswami2018indefinite,wei2019experimental}. An additional experiment recently built the quantum switch with more than two channels \cite{taddei2020experimental}.

Following the above investigations, the quantum switch was proposed as a new probing strategy for parameter estimation of quantum channels \cite{frey2019indefinite}. This strategy consists of placing copies of two or $N$ channels in indefinite causal order limited to only two different causal orders. Refs. \cite{zhao2020quantum,mukhopadhyay2018superposition} also showed a gain of the quantum Fisher information by using superposition of causal orders. However, the studies in Refs. \cite{zhao2020quantum,mukhopadhyay2018superposition} were also limited to two causal orders. An open question that remains is where there is additional gain in the quantum Fisher information when the number of causal orders is increased. In this work, we address this question by exploring the quantum 3-switch \cite{procopio2019communication} and performing fine control on selected combinations of causal orders. Such fine control was not accessible to the two-causal order case studied previously \cite{procopio2020sending}. We found that the quantum Fisher information increases as the number of causal orders increases for certain combinations of causal orders within specific regions of noise. We also found that certain combinations of causal orders are less efficient for increasing quantum Fisher information for the same number of causal orders. Based on our results, we chose the best combinations of causal orders with three channels for probing strategies using indefinite causal structures.

The structure of the paper is as follows.  In section~\ref{preliminaries}, we review the mathematical structure of the quantum switch with three noisy channels.  Then in section~\ref{Fisher} we review the basics of the quantum Fisher information and explain the procedure to calculate the quantum Fisher information for all combinations of alternatives orders with three channels.  Section \ref{results} presents the analytical expressions for the quantum Fisher information for the number of causal orders. Section \ref{discussion} presents the discussion of our results. Finally, Section \ref{concu} sums up our conclusions.

\section{The Quantum 3-Switch}\label{preliminaries}

In Ref. \cite{procopio2019communication}, it was introduced the quantum switch with three quantum channels  $\N{1}$, $\N{2}$ and $\N{3}$ can be in $\sum_{m=2}^{N!}\binom{N!}{m}$ superpositions of $m$ causal orders with $N=3$, where $\binom{n}{r}= \frac{n!}{r!(n-r)!}$ is the binomial coefficient.  A  noisy channel $\mathcal{N}$ on a qudit ($d$-dimensional) input state $\sigma$  can be modeled as 
\begin{equation} \label{depch2}
\mathcal{N}(\sigma) = \theta \sigma + (1-\theta) {\tr} [\sigma] \frac{\mathbb{1}}{d}, 
\end{equation}

\noindent where $\theta$ is the depolorazing parameter and  $\mathbb{1}$ is the identity operator, which represents a maximally mixed state.  We use the Kraus decomposition  $\mathcal{N}(\sigma)=\sum_{i} K_{i}  \sigma K_{i}^{\dagger}$ to mathematically represent the action of a channel ${\mathcal N}$ on the quantum state $\sigma$ such that $\sum_{i} K_{i}  K_{i}^{\dagger}=\mathbb{1}_t$. For three noisy channels $\mathcal{N}_1$, $\mathcal{N}_2$ and $\mathcal{N}_3$, the control system $\sigma_c=\left| \psi_c \right>  \left< \psi_c \right|  $  coherently controls  $\sigma$. The quantum state of the control system writes 
\begin{align} \label{Control3}
\left| \psi_c \right>
& = \sum_{k=1}^6\sqrt{ p_k} \left| k \right>,
\end{align}

\noindent   with $k \in [\![1;6!]\!]$ and $p_k$ is the probability associated one combination of orders as in  \cite{procopio2020sending} such that $ \sum_{k=1}^{6} p_k=1$.   Setting  the control state as in equation (\ref{Control3}), yields a superposition of several causal orders with their respective weights $p_k$. To select specific combinations of causal orders, one only needs to fix specific values to the  $p_k$. If the Kraus operators of the channels $\mathcal{N}_1$, $\mathcal{N}_2$ and $\mathcal{N}_3$ are $\{K_{i}^{(1)}\}$, $\{K_{j}^{(2)}\}$ and $\{K_{k}^{(3)}\}$ respectively, then the  Kraus operators $\mathcal{K}_{ijk}$ of the full quantum 3-switch channel is
\begin{align} \label{Kraus3}
\mathcal{K}_{ijk} = \sum_{k=1}^6 \pi_{k}( K_{i}^{(1)}K_{j}^{(2)}K_{k}^{(3)}) \left| k \right>\left< k \right|,
\end{align}

\noindent where $\pi_n $ is a permutation of the symmetric group $S_6=\{ \pi_k | k \in [\![1;6!]\!] \}$.  The action of the quantum $3$-switch  ${\mathcal S}(\N{1},\N{2},\N{3})$ over an input $\sigma \otimes \sigma_c $  can be identified as the output state $\rho_{{\rm out}}(\theta_1,\theta_2,\theta_3)=\mathcal S(\N{1},\N{2},\N{3}) \left( \sigma \otimes \sigma_c \right) $ of the quantum 3-switch and can be  expressed through the  Kraus operators $\mathcal{K}_{ijk}$ as
\begin{equation}\label{Krausg}
\rho_{{\rm out}}(\theta_1,\theta_2,\theta_3)=  \sum_{ijk} \mathcal{K}_{ijk}\left( \sigma \otimes \sigma_c \right.) \mathcal{K}_{ijk}^\dagger.
\end{equation}

\noindent Following the procedure described in Ref. \cite{procopio2019communication}, we found that the output of the  quantum 3-switch  is a   $6\times6$ block-symmetry matrix. For sake of simplicity, we study the one-parameter case, that is, we take three copies of the same  depolarizing channel $\mathcal{N}(\sigma)$, i.e. $\theta_1=\theta_2=\theta_3=\theta$, thus for this case the quantum 3-switch output can be written as
\begin{equation}\label{Q3S-2}
	\begin{array}{lll}
		{ \rho_{\rm out}^{(m)}}(\theta)=\frac{1}{m}\left(\begin{array}{cccccc} 
			\epsilon_{11}{A} &\epsilon_{12} {B}&\epsilon_{13}{B}&\epsilon_{14}{D}&\epsilon_{15}{D}&\epsilon_{16}{F}\\
			\epsilon_{12}{B} &\epsilon_{22}{A} &\epsilon_{23}{D}&\epsilon_{24}{F}&\epsilon_{25}{B}&\epsilon_{26}{D}\\ 
			\epsilon_{13}{B} &\epsilon_{23}{D} &\epsilon_{33}{A}&\epsilon_{34}{B}&\epsilon_{35}{F}& \epsilon_{36}{D}\\ 
			\epsilon_{14}{D} &\epsilon_{24}{F} & \epsilon_{34}{B}&\epsilon_{44} {A}&\epsilon_{45}{D}&\epsilon_{46}{B}\\
			\epsilon_{15}{D} &\epsilon_{25}{B}&\epsilon_{35} {F}&\epsilon_{45}{D}&\epsilon_{55}{A}& \epsilon_{56}{B}\\
			\epsilon_{16}{F}&\epsilon_{26}{D} & \epsilon_{36}{D}&\epsilon_{46}{B}& \epsilon_{56}{B}&\epsilon_{66}{A}\end{array}\right),
	\end{array}
\end{equation}

\noindent where we have introduced the parameter  $\epsilon_{ij}=m\sqrt{p_ip_j}$,  with $p_i=1/m$, such that  $0 \leq\epsilon_{ij}\leq 1 $, which reflects the degree of superposition between two definite causal orders determined by the probabilities $p_i$ and $p_j$, and the parameter  $m$ stands for  the number of causal orders involved in the superposition, with $m=1,\dots ,3!$. The matrices $A,B,D$ and $F$ from the block matrix (\ref{Q3S-2}) are linear combinations of the identity $\mathbb{1}$ and the input  state $\sigma$. These matrices can be written as
\begin{equation}\label{coefficients}
{X}= f_{X}(d,\theta) \mathbb{1}+ g_{X}(d, \theta) \sigma,
\end{equation}
\noindent where  $X \in \{{A},{D},{B},{F}\}$ and $f_{X}(d,\theta)$ and $g_{X}(d, \theta)$ are functions of the dimension $d$ and the depolarising parameter $\theta$, see Appendix A. Note that the matrices $X$ commute. This fact simplifies calculations as the matrices $X$ can be treated as matrix elements. Given that the input state $\sigma$  of the quantum 3-switch can have a spectral decomposition $\sigma=\sum_{i=1}^{d} \gamma_{i}^{\sigma} \left| i \right> \left< i \right|$, the matrices $X$ can have also a spectral decomposition as $X=\sum_{i=1}^{d} \gamma_{i}^{x} \left| i \right> \left< i \right|$, where the eigenvalue $\gamma_{i}^{x}$ for ${x}=\{{a},{d},{b},{f}\}$  corresponds to eigenvalues of matrices $A,B,D$ and $F$ respectively. These eigenvalues depend on  $\theta$ and  $d$.

\section{Quantum Fisher Information}\label{Fisher}

\noindent To find bounds on the precision of $\theta$  we use the Cramer-Rao inequality which establishes a lower bound on the variance Var of an estimator  $\hat{\theta}$
\begin{equation}\label{CR}
{\rm Var}(\hat{\theta})\geqslant\frac{1}{J(\theta)},
\end{equation}

\noindent where $J(\theta)$ is the quantum Fisher information defined as 
\begin{equation}\label{Jm}
J(\theta)=\tr[\rho(	\theta ) L^2],
\end{equation}
\noindent where the score operator $L$ is  the Hermitian symmetric logarithmic derivative (SLD) \cite{paris2009quantum} that satisfy the following equation
\begin{equation}\label{SLD}
2 \frac{\partial  \rho (\theta)}{\partial  \theta}=	L\rho(\theta) + \rho(\theta) L.
\end{equation}

\noindent The Cramer-Rao inequality tells us that the larger $J(\theta)$ is, the lower  Var$(\hat{\theta})$ will be. In other words, the precision to estimate the parameter $\theta$ is improved if the quantum Fisher information is larger. Thus, increasing the quantum Fisher information with different strategies is an important task in quantum metrology. Here we use the quantum 3-switch to improve the parameter estimation of $\theta$. For the  input state we consider an input $\sigma\otimes\sigma_c$ where $\sigma$ is considered  to be a pure state. The output state ${\rho_{\rm out}^{(m)}}(\theta) =\mathcal S(\N,\N,\mathcal{N} ) \left( \sigma \otimes \sigma_c \right)$ is the one-parameter family of density operators which are used to calculate the quantum Fisher information using equation (\ref{Jm}), i.e., we take $\rho(\theta)={\rho_{\rm out}^{(m)}}(\theta)$. It has been shown that the maximum quantum Fisher information is attained when the input state $\sigma$ is a pure sate \cite{fujiwara2005quantum}, that is,  $\sigma=\left| i \right> \left< i \right|$. Without loss of generality, we take the input state $\sigma=\left| 1 \right> \left< 1 \right|$. The identity $\mathbb{1}$ in equation (\ref{coefficients}) has a spectral decomposition as $\mathbb{1}=\sum_{i=1}^{d}  \left| i \right> \left< i \right|$ so that the matrices $X$ have the following spectral decomposition

\begin{equation}\label{Xdes}
{X}= (f_{X}(d,\theta)+g_{X}(d, \theta)) \left| 1 \right> \left< 1 \right| + \sum_{i=2}^{d} f_{X}(d,\theta) \left| i \right> \left< i \right| ,
\end{equation}

\noindent Comparing equation (\ref{Xdes}) with $X=\sum_{i=1}^{d} \gamma_{i}^{x} \left| i \right> \left< i \right|$ we found that the eigenvalues $\gamma_{i}^{x}$ of matrices $X$ are

\begin{equation}\label{eigenvalues}
\begin{array}{ll}
\gamma_{1}^{x}=f_{X}(d,\theta)+g_{X}(d, \theta) \vspace{2ex}, \forall  \hspace{0.5ex}  i=1,\\
\gamma_{i}^{x}=f_{X}(d,\theta) \hspace{2ex} \forall \hspace{0.5ex} i \geqslant 2.
\end{array}
\end{equation}

Our principal task is to determine the quantum Fisher information $J(\theta)$ using  equation (\ref{Jm}) for the one-parameter output channels  ${ \rho_{\rm out}^{(m)}}(\theta)$. To do that, we need first to specify the number of involved causal orders $m$   to define properly the structure  of the matrix (\ref{Q3S-2}). Once we specify ${ \rho_{\rm out}^{(m)}}(\theta)$ for a  given $m$, we calculate the corresponding  score operator $L$ from equation (\ref{SLD}). For a given $m$, the structure of the score operator  $L$ should have the same mathematical structure as ${ \rho_{\rm out}^{(m)}}(\theta)$. We specify this relation by having a superscript $m$ in $L$, i.e. $L^{(m)}$. We also specify the corresponding quantum Fisher information for a given $m$ as $J^{(m)}(\theta)$. The parameters $m$ and $\epsilon_{ij}$ in matrix (\ref{Q3S-2}) play an important role to select the desired superposition of causal orders.

The number of combinations to superimpose $m$ definite causal orders  with three channels is $\binom{3!}{m}$, where $\binom{n}{r}= \frac{n!}{r!(n-r)!}$ is the binomial coefficient. If we assign to each combination the corresponding output state ${ \rho_{\rm out}^{(m)}}(\theta)$,  we will have $\binom{3!}{m}$   different output states ${ \rho_{\rm out}^{(m)}}(\theta)$. In total, there will be $\sum_{m=2}^{3!}\binom{3!}{m}=57$ different output states ${ \rho_{\rm out}^{(m)}}(\theta)$ and therefore 57 different score operators $L^{(m)}$ to be calculated. However,  in Ref. \cite{procopio2020threefold}, it was found that the transmission of classical information with three channels in superposition of causal orders has a threefold behaviour. These behaviours are based on the properties of three equivalent classes of quantum switch matrices ${\rho_{\rm out}^{(m)}}(\theta)$.  For a given $m$, there are three different equivalent classes of matrices ${\rho_{\rm out}^{(m)}}(\theta)$ which have the same matrix invariants and eigenvalues. Based on this fact, we choose only one representative element ${\rho_{\rm out}^{(m)}}(\theta)$ of each classe to calculate the corresponding quantum Fisher information $J^{(m)}(\theta)$. This avoids to calculate the 57 different score operators $L^{(m)}$ since elements of ${\rho_{\rm out}^{(m)}}(\theta)$ of the same group have the same mathematical properties. In the next sections we calculate the representative operators $L^{(m)}$ of each group and then we obtain the corresponding quantum Fisher information $J^{(m)}(\theta)$ for each $m$. We also analyze the relative quantum Fisher information gain defined as
\begin{equation}\label{Jrel}
J_{\rm rel}^{(m)}(\theta)=\frac{J_{{}}^{(m)}(\theta)-J_{{\rm def}}^{}(\theta)}{J_{{\rm def}}(\theta)},
\end{equation}

\noindent where $J^{(m)}(\theta)$ is the quantum Fisher information associated to $\theta$  when the channels are in an indefinite causal order with $m$ alternative causal orders and $J_{{\rm def}}(\theta)$ is the quantum Fisher information for channels in a definite causal order.

\section{Quantum Fisher information for $m$ causal orders}\label{results}

Following the procedure described above, we derivate analytical expressions for the quantum Fisher information for each $m$ causal order.  

\textit{Causal order $m=2$}. For this case, for the all combinations with $m=2$, we have only four nonzero matrix elements in ${\rho_{\rm out}^{(2)}}(\theta)$ and the rest of elements are zero. Here and for the next cases, we solve the problem to find $L^{(m)}$ with the corresponding  matrices ${\rho_{\rm out}^{(m)}}(\theta)$, but to avoid to write  $6\times6$ matrices with most of its elements zeros, we write ${\rho_{\rm out}^{(m)}}(\theta)$ as  $m\times m$ matrices. The problem to calculate $L^{(m)}$ should be equivalent using both type of matrices. The representative  matrices ${\rho_{\rm out}^{(2)}}(\theta)$ of each group  can be written as
\begin{equation}\label{Sol-m2}
	\begin{array}{lll}
		{\rho}_{X}^{(2)}(\theta)= \left(
		\begin{array}{cc}
			A p_1 &   \frac{1}{2}\epsilon_{12} X \\
		  \frac{1}{2}\epsilon_{12} X &  A p_2  \\
		\end{array}
		\right),
		\end{array}
\end{equation}

\noindent that is,  there are three different quantum switch matrices, ${\rho}_{B}^{(2)}(\theta)$, ${\rho}_{D}^{(2)}(\theta)$  and  $	{\rho}_{F}^{(2)}(\theta)$ which represent the three equivalent classes of quantum 3-switch matrices that can be found with two causal orders in superposition.  Given the block form of matrices (\ref{Sol-m2}), the score operator $L_{}^{(2)}$ has also  three different score operators ${ L}_{B}^{\scaleto{(2)}{5pt}}, { L}_{D}^{\scaleto{(2)}{5pt}}$ and ${ L}_{F}^{\scaleto{(2)}{5pt}}$  with a corresponding block form 
\begin{equation}\label{Score-m2}
	\begin{array}{ll}
		L_{}^{(2)}=\left(
		\begin{array}{cc}
			L_{11}  & L_{12}  \\
			L_{12} & L_{22} \\
		\end{array}
		\right).
	\end{array}
\end{equation}

\noindent  By introducing the output states (\ref{Sol-m2}) and the score operator (\ref{Score-m2}) into (\ref{SLD}), we have the following equation systems
\begin{equation}
\begin{array}{ll}
2 A L_{11} p_1+B L_{12} \epsilon _{12}=2 p_1 \partial_\theta A,
  \vspace{3ex}\\
2 A L_{22} p_2+X L_{12} \epsilon _{12}=2 p_2 \partial_\theta A, \vspace{3ex}\\
A L_{12} +\frac{\epsilon _{12}}{2} X L_{11} +\frac{\epsilon _{12}}{2} X L_{22} =\epsilon _{12} \partial_\theta X , \vspace{3ex}\\
\end{array}
\end{equation}
\noindent whose solutions are
\begin{equation}\label{solutionsm2}
	\begin{array}{ll}
		L_{11}=\text{$A$}^{-1}(\partial_{\theta}\text{$A$}-2p_1\text{$X$} \Phi_{\text{$X$}}),  \vspace{3ex}\\
		L_{22}=\text{$A$}^{-1}(\partial_{\theta}\text{$A$}-2p_j\text{$X$} \Phi_{\text{$X$}})  \vspace{3ex} , \\%
		L_{12}= \epsilon_{1j} \Phi_{\text{$X$}} , \vspace{3ex}\\%
	\end{array}
\end{equation}
\noindent where $\Phi_X=\frac{{A} \partial_\theta \text{${X}$}- {X}\partial_\theta\text{${A}$} }{{A}^2-{X}^2}$. By introducing solutions (\ref{solutionsm2}) into (\ref{Jm}) we found that the quantum Fisher information is 
\begin{equation}\label{Jm2}
J^{(2)}_X(d,\theta)=\tr[A^{-1}(\partial_{\theta}A)^2]+ \epsilon _{1j} \tr[A^{-1}(A^2-X^2)\Phi^2_X)],
\end{equation}

\noindent where the term $\tr[A^{-1}(\partial_{\theta}A)^2]=\sum_{i=1}^{d}\frac{(\partial_{\theta} \gamma_i^a)^2}{\gamma_i^a}$ in (\ref{Jm2}) is the quantum Fisher information associated with $\theta$ when the three channels are in a definite order, that is
\begin{equation}
J^{{\rm N3}}_{{\rm def}}(d,\theta)\equiv\tr[A^{-1}(\partial_{\theta}A)^2]=\frac{9 (d-1) \theta^4}{\left(1-\theta^3\right) \left((d-1) \theta^3+1\right)}.
\end{equation}

\noindent The second term in equation (\ref{Jm2}) is associated with $\theta$ when the three channels are in an indefinite order 
\begin{equation}\label{secondm2}
J^{(2)}_{{{\rm ind,X}}}(d,\theta)\equiv \tr[A^{-1}(A^2-X^2)\Phi^2_X)]=\sum_{i=1}^{d}\frac{(\gamma_{i}^a\partial_\theta \gamma_i^x -\gamma_{i}^x\partial_\theta \gamma_i^a )^2}{\gamma_i^a((\gamma_{i}^a)^2 -(\gamma_{i}^x)^2 )}.
\end{equation}

\begin{figure}\label{Fishers}
	\begin{center} 
		\scalebox{.55}{\includegraphics[width = .8\textwidth]{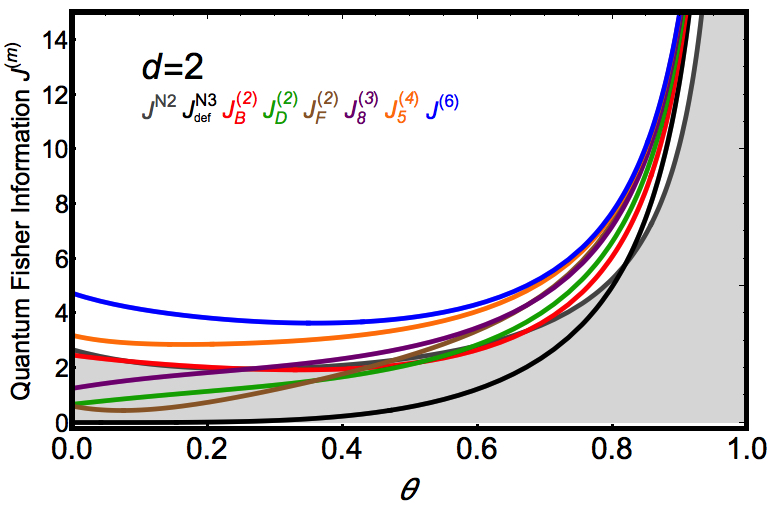}} \textbf{(a)}
		\scalebox{.55}{\includegraphics[width = .8\textwidth]{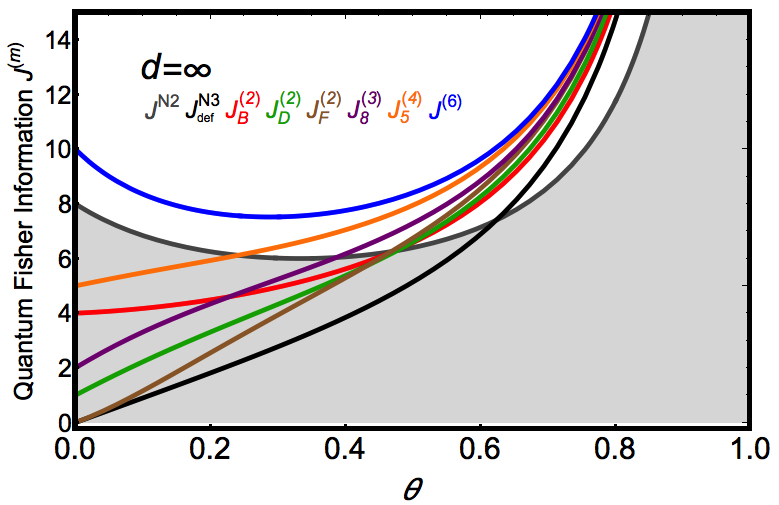}} \textbf{(b)}
		\scalebox{.55}{\includegraphics[width = .8\textwidth]{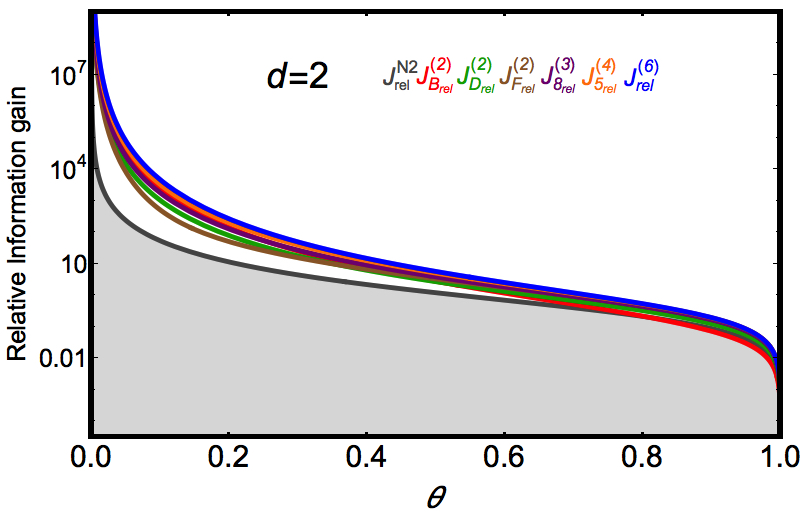}} \textbf{(c)}
		\scalebox{.55}{\includegraphics[width = .8\textwidth]{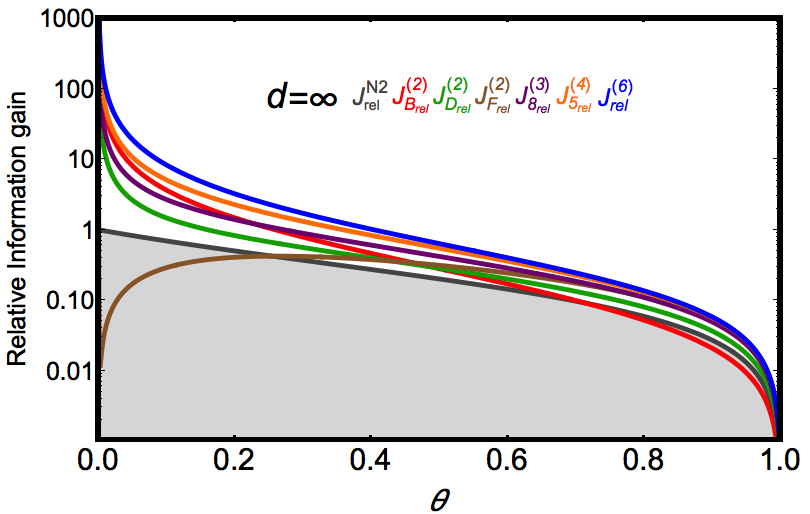}}\textbf{(d)}
		\caption{\footnotesize \textbf{ Comparison between  the quantum Fisher informations of  different number of causal orders $J^{(m)}$}.  Figures (a) and (b) show the quantum Fisher information $J^{(m)}$ for different $m$ causal orders as a function of the depolorazing parameter  $\theta$ for two  different dimensions of the input state $\sigma$, $d=2$ and $d\to \infty$  respectively.     Figures (c) and (d) show the relative quantum Fisher gain $J^{(m)}_{\rm rel}$ for dimensions $d=2$ and $d\to \infty$ respectively.    $ J^{\rm N3}_{\rm def}$ is the quantum Fisher information of  three channels in definite causal order.  The gray zone is for delimiting the region of the quantum Fisher information $ J^{\rm N2}$ of the quantum 2-switch. $ J^{\rm N2}$ was calculated from Ref. \cite{frey2019indefinite}.}
		\label{JXm2figure}
	\end{center}
\end{figure}

\noindent  By substituting the eigenvalues  (\ref{eigenvalues}) of matrices $B, D$ and $F$ in (\ref{secondm2}),  we can calculate the quantum Fisher information (\ref{Jm2}) as a function of the parameters $d$ and $\theta$. Equations (\ref{Jm2Bexp}), (\ref{Jm2Dexp}) and (\ref{Jm2Fexp}) from Appendix B show the corresponding quantum Fisher information expressions $J^{(2)}_{{{\rm ind,X}}}(d,\theta)$. Figure \ref{JXm2figure} shows the quantum Fisher information as a function of $\theta$ and $d$.  For complete depolarizing channels, i.e. $\theta$=0,  in the limit when $d\to \infty$, we  have  $J^{(2)}_B(d)>J^{(2)}_D(d)>J^{(2)}_F(d)$.\\

\noindent \textit{Causal order $m=3$}. For this case we have also three equivalent classes of quantum 3-switch. We also expect to have  three different behaviours for the quantum Fisher information as we found in the case of $m=2$  causal orders. However, by calculating the score operators from the system of equations given by equation (\ref{Jm}), only the  representative element 
\begin{equation}\label{Q3S-m3}
\begin{array}{lll}
 {\rho}_{8}^{\scaleto{(3)}{5pt}}(\theta)=\frac{1}{3}\left(
\begin{array}{cccccc}
A & D & D  \\
D  & A & D  \\
D  & D & A  \\
\end{array}
\right),
\end{array}
\end{equation}

\noindent  can be used to calculate readily  the score operator 
\begin{equation}\label{Score-m3}
{L}_{8}^{\scaleto{(3)}{5pt}}=\left(
\begin{array}{cccccc}
{L}_{}^{\scaleto{(3)}{5pt}}  & {L}_{D}^{\scaleto{(3)}{5pt}} &{L}_{D}^{\scaleto{(3)}{5pt}}  \\
{L}_{D}^{\scaleto{(3)}{5pt}}  & {L}_{}^{\scaleto{(3)}{5pt}} & {L}_{D}^{\scaleto{(3)}{5pt}}  \\
{L}_{D}^{\scaleto{(3)}{5pt}}  & {L}_{D}^{\scaleto{(3)}{5pt}} & {L}_{}^{\scaleto{(3)}{5pt}}  \\
\end{array}
\right),
\end{equation}

\noindent where the subscript 8 stands for the label of  the quantum 3-switch matrix associated to a specific combination of causal order, see \cite{procopio2020threefold}.  By introducing the output state (\ref{Q3S-m3}) and the score operator (\ref{Score-m3}) into (\ref{SLD}), we found that the block matrix elements from (\ref{Score-m3}) are 
\begin{equation}\label{L3}
{L}_{}^{\scaleto{(3)}{5pt}} =\frac{A \partial_{\theta} A+D \partial_{\theta} A-2 D\partial_{\theta} D  }{A^2+AD-2 D^2},
\end{equation}

\begin{equation}\label{L3D}
{L}_{D}^{\scaleto{(3)}{5pt}} =\frac{A \partial_{\theta} D-D \partial_{\theta} A  }{A^2+AD-2 D^2}.
\end{equation}

\noindent Substituting equations (\ref{L3}) and (\ref{L3D}) into (\ref{Jm}), we found that  the quantum Fisher information for $m=3$ causal orders is
\begin{equation} \label{Jm38}
J^{(3)}_{8}(d,\theta)={\rm Tr} \left[ \frac{ (\partial_{\theta} A)^2+2 (\partial_{\theta} D)^2}{(A-D) (A+2 D)} A \right] + {\rm Tr} \left[\frac{ \partial_{\theta} A^2-4 \partial_{\theta} A \partial_{\theta} D}{(A-D) (A+2 D)} D \right].
\end{equation}

\noindent By substituting the eigenvalues  (\ref{eigenvalues}) of matrices $A$ and $D$ in (\ref{Jm38}), we  obtain the quantum Fisher information in terms of $\theta$ and $d$, see equation (\ref{Jm38exp}) from Appendix B.\\

\noindent \textit{Causal order $m=4$}. For this case, we only calculate the quantum Fisher information for the following representative element
\begin{equation}\label{Q3S-m4}
\begin{array}{lll}
{\rho}_{5}^{\scaleto{(4)}{5pt}}(\theta)=\frac{1}{4}\left(
\begin{array}{cccccc}
A & B  & D  & F \\
B & A & F  & D \\
D & F  & A & B \\
F & D  & B  & A \\
\end{array}
\right).
\end{array}
\end{equation}
\noindent Likewise, given the block form of matrix (\ref{Q3S-m4}), the score operator ${ L}_{}^{\scaleto{(4)}{5pt}}$ has the corresponding block form 
\begin{equation} \label{Score-m4}
{L}_{5}^{\scaleto{(4)}{5pt}}=\left(
\begin{array}{cccccc}
{L}_{}^{\scaleto{(4)}{5pt}}  & {L}_{B}^{\scaleto{(4)}{5pt}}   & {L}_{D}^{\scaleto{(4)}{5pt}}   & {L}_{F}^{\scaleto{(4)}{5pt}}  \\
{L}_{B}^{\scaleto{(4)}{5pt}}  &{L}_{}^{\scaleto{(4)}{5pt}}  & {L}_{F}^{\scaleto{(4)}{5pt}}   & {L}_{D}^{\scaleto{(4)}{5pt}}  \\
{L}_{D}^{\scaleto{(4)}{5pt}}  & {L}_{F}^{\scaleto{(4)}{5pt}}  & {L}_{}^{\scaleto{(4)}{5pt}}  & {L}_{B}^{\scaleto{(4)}{5pt}}  \\
{L}_{F}^{\scaleto{(4)}{5pt}}  & {L}_{D}^{\scaleto{(4)}{5pt}}   & {L}_{B}^{\scaleto{(4)}{5pt}}   & {L}_{}^{\scaleto{(4)}{5pt}} \\
\end{array}
\right).
\end{equation}

\noindent By introducing the output states (\ref{Q3S-m4}) and the score operator (\ref{Score-m4}) into (\ref{SLD}), we found that 
\begin{equation}
\begin{array}{ll}
{L}_{}^{\scaleto{(4)}{5pt}}=\frac{1}{4} \left(\frac{\partial_{\theta} \Phi _{\alpha }}{\Phi _{\alpha }}+\frac{\partial_{\theta} \Phi _{\beta }}{\Phi _{\beta }}+\frac{\partial_{\theta} \Phi _{\gamma }}{\Phi _{\gamma }}+\frac{\partial_{\theta} \Phi _{\sigma }}{\Phi _{\sigma }}\right), \hspace{3ex}
 {L}_{B}^{\scaleto{(4)}{5pt}}=\frac{1}{4} \left(\frac{\partial_{\theta} \Phi _{\alpha }}{\Phi _{\alpha }}+\frac{\partial_{\theta} \Phi _{\beta }}{\Phi _{\beta }}-\frac{\partial_{\theta} \Phi _{\gamma }}{\Phi _{\gamma }}-\frac{\partial_{\theta} \Phi _{\sigma }}{\Phi _{\sigma }}\right),\vspace{3ex}\\ 
 {L}_{D}^{\scaleto{(4)}{5pt}}=\frac{1}{4} \left(-\frac{\partial_{\theta} \Phi _{\alpha }}{\Phi _{\alpha }}+\frac{\partial_{\theta} \Phi _{\beta }}{\Phi _{\beta }}-\frac{\partial_{\theta} \Phi _{\gamma }}{\Phi _{\gamma }}+\frac{\partial_{\theta} \Phi _{\sigma }}{\Phi _{\sigma }}\right),\hspace{1ex}
{L}_{F}^{\scaleto{(4)}{5pt}}=\frac{1}{4} \left(-\frac{\partial_{\theta} \Phi _{\alpha }}{\Phi _{\alpha }}+\frac{\partial_{\theta} \Phi _{\beta }}{\Phi _{\beta }}+\frac{\partial_{\theta} \Phi _{\gamma }}{\Phi _{\gamma }}-\frac{\partial_{\theta} \Phi _{\sigma }}{\Phi _{\sigma }}\right),
\end{array}
\end{equation}
\noindent where $\Phi$'s are linear combinations of matrices $A,B,D$ and $F$
\begin{equation}
\begin{array}{ll}
\Phi _{\alpha }=A+B-D-F,\\
\Phi _{\beta }=A+B+D+F,\\ 
\Phi _{\gamma }=A-B-D+F,\\
\Phi _{\sigma }=A-B+D-F.
\end{array}
\end{equation}
For this case, we found that the quantum Fisher information (\ref{Jm})  can be  calculated as 
\begin{equation} \label{Jm45}
\begin{array}{ll}
J^{(4)}_{5}(d,\theta)=\frac{1}{4} \left(\text{Tr}\left[(\Phi _{\alpha })^{-1}\left(\partial_{\theta} \Phi _{\alpha }\right)^2\right]+\text{Tr}\left[(\Phi _{\beta })^{-1}\left(\partial_{\theta} \Phi _{\beta }\right)^2 \right] + \right.  
\vspace{3ex}  \\ \hspace{12ex}
\left. \text{Tr}\left[(\Phi _{\gamma })^{-1}\left(\partial_{\theta} \Phi _{\gamma }\right)^2 \right]+\text{Tr}\left[(\Phi _{\sigma })^{-1}\left(\partial_{\theta} \Phi _{\sigma }\right)^2 \right]\right).
\end{array}
\end{equation}

\noindent The corresponding analytical expression of (\ref{Jm45}) in terms of $\theta$ and $d$ can be found in equation (\ref{Jm45exp}) from Appendix B.\\

\noindent \textit{Causal order $m=5$}. For this case,  we were unable to calculate directly the components of the score operators using the system of equations given by (\ref{SLD}). Other methods can be used to determine the score operator, for example see reference \cite{vsafranek2018simple}.\\

\noindent \textit{Causal order $m=6$}. For this case, we are able to calculate directly the score operator from the system of equations  given by  (\ref{SLD}). The quantum 3-switch output by taking account all the six alternative causal orders is 
\begin{equation}\label{Q3S-m6}
\begin{array}{lll}
{\rho}^{\scaleto{(6)}{5pt}}(\theta)= \frac{1}{6}\left(\begin{array}{cccccc} 
A &B & B&D&D&F\\
B &A &D&F&B&D\\ 
B &D & A&B&F&D\\ 
D &F & B&A&D&B\\
D &B& F&D&A&B\\
F&D & D&B&B&A\end{array}\right),
\end{array}
\end{equation}

\noindent Likewise, given the block form (\ref{Q3S-m6}), the score operator ${ L}_{}^{\scaleto{(6)}{5pt}}$ has the corresponding block form 

\begin{equation}\label{Score-m6}
L^{\scaleto{(6)}{5pt}}=\left(
\begin{array}{cccccc}
L^{\scaleto{(6)}{5pt}} & L_B^{\scaleto{(6)}{5pt}} & L_B^{\scaleto{(6)}{5pt}} & L_D^{\scaleto{(6)}{5pt}} & L_D^{\scaleto{(6)}{5pt}} & L_F^{\scaleto{(6)}{5pt}} \\
L_B^{\scaleto{(6)}{5pt}} & L^{\scaleto{(6)}{5pt}} & L_D^{\scaleto{(6)}{5pt}} & L_F^{\scaleto{(6)}{5pt}} & L_B^{\scaleto{(6)}{5pt}} & L_D^{\scaleto{(6)}{5pt}} \\
L_B^{\scaleto{(6)}{5pt}} & L_D & L^{\scaleto{(6)}{5pt}} & L_B^{\scaleto{(6)}{5pt}} & L_F^{\scaleto{(6)}{5pt}} & L_D^{\scaleto{(6)}{5pt}} \\
L_D & L_F^{\scaleto{(6)}{5pt}} & L_B^{\scaleto{(6)}{5pt}} & L^{\scaleto{(6)}{5pt}} & L_D^{\scaleto{(6)}{5pt}} & L_B^{\scaleto{(6)}{5pt}} \\
L_D & L_B^{\scaleto{(6)}{5pt}} & L_F^{\scaleto{(6)}{5pt}} & L_D^{\scaleto{(6)}{5pt}} & L^{\scaleto{(6)}{5pt}} & L_B^{\scaleto{(6)}{5pt}} \\
L_F^{\scaleto{(6)}{5pt}} & L_D^{\scaleto{(6)}{5pt}} & L_D^{\scaleto{(6)}{5pt}} & L_B^{\scaleto{(6)}{5pt}} & L_B^{\scaleto{(6)}{5pt}} & L^{\scaleto{(6)}{5pt}} 
\end{array}
\right).
\end{equation}

\noindent By introducing the output states (\ref{Q3S-m6}) and  (\ref{Score-m6}) into (\ref{SLD}), we found that

\begin{equation}
\begin{array}{ll}
{L}_{}^{\scaleto{(6)}{5pt}}=\frac{1}{6} \left(2 \frac{\partial_{\theta} \Phi _{\alpha }}{\Phi _{\alpha }}+2\frac{\partial_{\theta} \Phi _{\gamma }}{\Phi _{\gamma }}+\frac{\partial_{\theta} \Phi _{\mu }}{\Phi _{\mu }}+\frac{\partial_{\theta} \Phi _{\eta }}{\Phi _{\eta }}\right), \hspace{3ex}
{L}_{B}^{\scaleto{(6)}{5pt}}=\frac{1}{6} \left(\frac{\partial_{\theta} \Phi _{\alpha }}{\Phi _{\alpha }}-\frac{\partial_{\theta} \Phi _{\gamma }}{\Phi _{\gamma }}-\frac{\partial_{\theta} \Phi _{\mu }}{\Phi _{\mu }}+\frac{\partial_{\theta} \Phi _{\eta }}{\Phi _{\eta }}\right),\hspace{3ex} \vspace{3ex}\\  
{L}_{D}^{\scaleto{(6)}{5pt}}=\frac{1}{6} \left(-\frac{\partial_{\theta} \Phi _{\alpha }}{\Phi _{\alpha }}-\frac{\partial_{\theta} \Phi _{\gamma }}{\Phi _{\gamma }}+\frac{\partial_{\theta} \Phi _{\mu }}{\Phi _{\mu }}+\frac{\partial_{\theta} \Phi _{\eta }}{\Phi _{\eta }}\right),\hspace{3ex}
{L}_{F}^{\scaleto{(6)}{5pt}}=\frac{1}{6} \left(-2\frac{\partial_{\theta} \Phi _{\alpha }}{\Phi _{\alpha }}+ 2\frac{\partial_{\theta} \Phi _{\gamma }}{\Phi _{\gamma }}-\frac{\partial_{\theta} \Phi _{\mu }}{\Phi _{\mu }}+\frac{\partial_{\theta} \Phi _{\eta }}{\Phi _{\eta }}\right),
\end{array}
\end{equation}

\noindent where 
\begin{equation}
\begin{array}{ll}
\Phi _{\alpha }=A+B-D-F,\\
\Phi _{\eta }=A+2B+2D+F,\\
\Phi _{\gamma }=A-B-D+F,\\
\Phi _{\mu }=A-2B+2D-F.
\end{array}
\end{equation}

\noindent Thus, taking account all causal orders with three channels we found that the quantum Fisher information for $m=6$  is 
\begin{equation} \label{Jm6}
\begin{array}{ll}
J^{(6)}_{}(d,\theta)=\frac{1}{3} \text{Tr}\left[(\Phi _{\alpha })^{-1}\left(\partial_{\theta} \Phi _{\alpha }\right)^2\right]+ \frac{1}{6}\text{Tr}\left[(\Phi _{\eta })^{-1}\left(\partial_{\theta} \Phi _{\eta }\right)^2 \right] +\\
\hspace{12ex} \frac{1}{3}\text{Tr}\left[(\Phi _{\gamma })^{-1}\left(\partial_{\theta} \Phi _{\gamma }\right)^2 \right]+\frac{1}{6}\text{Tr}\left[(\Phi _{\mu })^{-1}\left(\partial_{\theta} \Phi _{\mu }\right)^2 \right],
\end{array}
\end{equation}
	
\noindent whose analytical expression of (\ref{Jm6}) in terms of $\theta$ and $d$ can be found in equation (\ref{Jm6exp}) from Appendix B. 

\section{Discussion}\label{discussion}

In Figure~\ref{Fishers}, we compare the quantum Fisher information with $m$ causal orders as a function of the noise parameter $\theta$. As you can see in Figure 1(a), for certain combinations of causal orders, the quantum Fisher information is lower for the quantum 3-switch than the quantum 2-switch in the noise region $0<\theta<0.5$. We also see that the quantum Fisher information $J^{(4)}_{5}$ is always higher than the quantum Fisher information for the two-case scenario in any noise region. This allows us to choose the best combination of causal orders with three channels to minimize the resources needed to implement the quantum switch, and to exceed the value of the quantum Fisher information of the two-channel scenario. Figure 1(a) also shows that $J^{(6)}_{}$  is higher than any quantum Fisher information of different combination of causal orders. This confirms the fact that if we take all the alternatives of causal orders in superposition, we will have the maximum gain in the quantum Fisher information. In addition, the quantum Fisher information with three channels $J^{(6)}_{}$  is always higher than the quantum Fisher information $J^{N2}_{}$ for two channels.  This may well indicate that the quantum Fisher information increases as the number of channels increases.  However, in Ref. \cite{chiribella2019quantum}, it has been shown that the classical capacity increases monotonically with the number of channels, but it asymptotically reaches a limit for larger number of channels. The question whether the quantum Fisher information increases or has a limit as the number of channels increases is a fundamental question that future investigations could be done. In Figure 1(b), as the dimension $d$ of the input state $\sigma$ tends to infinity, the quantum Fisher information for $m<6$ causal orders is lower than the quantum Fisher information for the quantum 2-switch in the noisy regions $0<\theta<0.3$. Remarkably,  $J^{(6)}_{}$  is always higher than the two-channel scenario at large dimensions of $d$ for any noise region. Figures 1(c)-1(d) show the relative quantum Fisher information gain  defined in equation (\ref{Jrel}). Finally, we notice that the quantum Fisher information $J^{(6)}_{}$  was also calculated in Ref. \cite{frey2021quantum} following a different procedure. Ref. \cite{frey2021quantum} only studies three combinations among the 57 possible combinations to superimpose different causal orders. Our procedure allowed us to study the Fisher information for any combination of causal orders through the quantum switch matrix (\ref{Q3S-2}) and to provide analytical and compact expressions for the quantum Fisher information.

\section{Conclusions}\label{concu}

We studied the quantum Fisher information for any combination of causal orders with three channels. Our procedure allows us to calculate the quantum Fisher information in closed analytical expressions for any causal order. We found that the quantum Fisher information increases as the number of causal orders increases within specific regions of noise and with specific combinations of causal orders. This allows us to choose the best indefinite causal structures to obtain the maximum possible values of the quantum Fisher information with less experimental resources. Our work can be extended to further study indefinite causal structures for estimating the value of parameters of other quantum channels of communications.

\ack
The author acknowledges the support of Israel Science Foundation and thanks  Nadia Belabas, Francisco Delgado and Alastair Abbott for discussions and valuable comments on this manuscript. 

\appendix

\section{Functions $f_X$ and $g_X$}\label{functionsX}
\vspace{1ex}

\noindent The functions $f_X$ and $g_X$ are the coefficients of equation (\ref{coefficients}):
\begin{equation}
\begin{array}{ll}
f_{A}(d,\theta)=\frac{1}{d} (1-\theta^3), \hspace{3ex} g_{A}(d, \theta)=\theta^3 \vspace{3ex},\\%
f_{D}(d,\theta)=\frac{1}{d} \left(-2\theta^3+\theta^2+\theta\right),\hspace{3ex}  g_{D}(d, \theta)=\frac{1}{d^2} \left(\left(d^2+1\right)\theta^3-\theta^2-\theta+1\right),   \hspace{3ex} \vspace{3ex}\\ 
f_{B}(d,\theta)= -\frac{1 }{d^3}(\theta-1)\left(\left(d^2+1\right) \theta^2+2 \left(d^2-1\right) \theta+1\right),\hspace{3ex}      g_{B}(d, \theta)=\frac{\theta  }{d^2} \left(\left(d^2+1\right) \theta^2-2 \theta+1\right),	 \vspace{3ex}\\%
f_{F}(d,\theta)=\frac{1}{d^3} \left((1-\theta)^3-3 d^2 (\theta-1)\theta^2\right) ,\hspace{3ex}      g_{F}(d, \theta)=\frac{1}{d^2} \left(d^2 \theta^3+3 (\theta-1)^2 \theta\right),	\vspace{3ex}\\%
\end{array}
\end{equation}

\noindent  where $\theta$ is the depolorazing parameter and $d$ the dimension of the input state $\sigma$.

\section{Analytical expressions  of the quantum Fisher Information} \label{Expressions}
\vspace{1ex}
\noindent In this section we present explicitly the expressions for the quantum Fisher information  $J^{(m)}_{{{}}}$  for $m$ causal orders. We start presenting the case where there are $m=2$ causal orders:   

\begin{equation} \label{Jm2Bexp}
\begin{array}{ll}
J^{(2)}_{{{\rm ind,B}}}(d,\theta)=
\frac{(d-1)}{d} \left(\frac{\left(d^2 \left(\theta^2-2 \theta-2\right)-3 \theta^2+3\right)^2}{\left(\theta^2+\theta+1\right) \left(d^4 \left(2 \theta^2+3 \theta+1\right)+2 d^2 \theta \left(\theta^2+\theta-2\right)+(\theta-1)^3\right)} \right.  + \vspace{3ex}  \\ \hspace{12ex}
\left. \frac{\left(d^2 (\theta-3) \theta^2+2 d \left(\theta^3-1\right)-3 (\theta+1) (\theta-1)^2\right)^2}{\left((d-1) \theta^3+1\right) \left(2 d^4 \theta^3-d^3 \left(2 \theta^4+\theta^2-2 \theta-1\right)+d^2 \left(2 \theta^4-5 \theta^2+2 \theta+1\right)-d (\theta-1)^3 (\theta+1)+(\theta-1)^4\right)}\right),
\end{array}
\end{equation}

\begin{equation}\label{Jm2Dexp}
\begin{array}{ll}
J^{(2)}_{{{\rm ind,D}}}(d,\theta)=\frac{(d-1)}{d} \left(\frac{\left((d-1) \theta^3+3 (d-1) \theta^2+3 \theta+1\right)^2}{(\theta+1) \left((d-1) \theta^3+1\right) \left(2 d^2 \theta^3+d \left(-3 \theta^3+\theta^2+\theta+1\right)+(\theta-1)^2 (\theta+1)\right)} \right.  +
\vspace{3ex}  \\ \hspace{19ex}
\left. \frac{\left(\theta^2+4 \theta+1\right)^2}{\left(\theta^2+\theta+1\right) \left(3 \theta^3+5 \theta^2+3 \theta+1\right)}\right),
\end{array}
\end{equation}

\begin{equation}\label{Jm2Fexp}
\begin{array}{ll}
J^{(2)}_{{{\rm ind,F}}}(d,\theta)=\frac{9 (d-1)}{d^7} \left(\frac{(d-1) \left(\theta^2-1\right)^2 ((d-1) \theta+1)^4}{\left((d-1) \theta^3+1\right) \left(\frac{\left((d-1) \theta^3+1\right)^2}{d^2}-\frac{((d-1) \theta+1)^6}{d^6}\right)}  \right.  +
\vspace{3ex}  \\ \hspace{19ex}
\left.  \frac{d^6 \left(d^2 \theta (\theta+2)+\theta^2-1\right)^2}{\left(\theta^2+\theta+1\right) \left(d^4 \left(8 \theta^3+6 \theta^2+3 \theta+1\right)+6 d^2 (\theta-1) \theta^2+(\theta-1)^3\right)}\right),
\end{array}
\end{equation}

\noindent where the labels $B,D$ and $F$ correspond to the block matrix elements from (\ref{Q3S-2}). Likewise, $\theta$ is the depolorazing parameter and $d$ the dimension of the input state $\sigma$. Each expression correspond to the evaluation of quantum Fisher information (\ref{Jm2})  for each equivalent class of the quantum switch matrices with $m=2$ causal orders.  For $m=3$ causal orders, the complete expression for the quantum Fisher information (\ref{Jm38}) is

\begin{equation}\label{Jm38exp}
\begin{array}{ll}
J^{(3)}_{8}(d,\theta)=\frac{(d-1) \left(3 d^2 \left(2 \theta^3-33 \theta^2-21 \theta-2\right) \theta^3+2 d (3 \theta+1)^2 \left(5 \theta^3-2 \theta^2-2 \theta-1\right)-6 \left(\theta^3+3 \theta^2-3 \theta-1\right)^2\right)}{d (\theta-1) (\theta+1) \left(5 \theta^2+3 \theta+1\right) \left(3 d^2 \theta^3+d \left(-5 \theta^3+2 \theta^2+2 \theta+1\right)+2 (\theta-1)^2 (\theta+1)\right)}.
\end{array}
\end{equation}

\noindent where the subscript 8 stands for the label of  the quantum 3-switch matrix associated to a specific combination of causal order, see \cite{procopio2020threefold}. For $m=4$ causal orders, the complete expression for the quantum Fisher information (\ref{Jm45}) is

\begin{equation} \label{Jm45exp}
\begin{array}{ll}
J^{(4)}_{5}(d,\theta)=\frac{1}{4} (d-1) \left[\frac{(3 d \theta+d-6 \theta+6)^2}{d^3 (d \theta+d-2 \theta+2)}+\frac{(1-9 \theta)^2}{d^2 (3 \theta+1)}+\frac{9-9 \theta}{d^2}-\frac{9 \left(d^2 \left(-7 \theta^2+2 \theta+1\right)-2 (\theta-1)^2\right)^2}{d^3 (\theta-1) \left(d^2 \left(7 \theta^2+4 \theta+1\right)+2 (\theta-1)^2\right)}
\right. +  \vspace{3ex}  \\ \hspace{19ex}
\left. \frac{9 (d-1) \left(4 d^2 \theta^2+d \left(-3 \theta^2+2 \theta+1\right)+2 (\theta-1)^2\right)^2}{d^3 \left(4 d^3 \theta^3+d^2 \left(-7 \theta^3+3 \theta^2+3 \theta+1\right)+d (5 \theta+1) (\theta-1)^2-2 (\theta-1)^3\right)} \right.  +
\vspace{3ex}  \\ \hspace{19ex}
\left.  \frac{\left(3 d^2 \theta+d^2+6 \theta-6\right)^2}{d^3 \left(d^2 (\theta+1)+2 (\theta-1)\right)}+\frac{(1-9 \theta)^2}{3 d \theta+d}+\frac{9-9 \theta}{d} \right].
\end{array}
\end{equation}

\noindent Finally, for $m=6$ causal orders, the expression for the quantum Fisher information (\ref{Jm6}) is

\begin{equation} \label{Jm6exp}
\begin{array}{ll}
J^{(6)}_{}(d,\theta)=\frac{( d-1)}{6 d^3}  \left[ -\frac{\left(9 (\theta-1)^2-6 d^2 \left(-5 \theta^2+\theta+1\right)\right)^2}{(\theta-1) \left(d^2 \left(10 \theta^2+7 \theta+1\right)+3 (\theta-1)^2\right)}+\frac{\left(2 d^2+9 \theta-9\right)^2}{d^2+3 \theta-3}\right.  +
\vspace{3ex}  \\ \hspace{19ex}
\left. \frac{9 (d-1) \left(6 d^2 \theta^2+d \left(-4 \theta^2+2 \theta+2\right)+3 (\theta-1)^2\right)^2}{6 d^3 \theta^3+d^2 \left(-10 \theta^3+3 \theta^2+6 \theta+1\right)+d (7 \theta+2) (\theta-1)^2-3 (\theta-1)^3} \right. + \vspace{3ex}  \\ \hspace{19ex} 
\left.  \frac{2 d (d+1) (1-9 \theta)^2}{3 \theta+1}+\frac{(2 d-9 \theta+9)^2}{d-3 \theta+3}-18 d (d+1) (\theta-1) \right].
\end{array}
\end{equation}

\end{document}